\documentclass[english,prd,twocolumn,superscriptaddress,nofootinbib,preprintnumbers]{revtex4-1}
\usepackage[latin1]{inputenc}
\usepackage{graphicx}
\usepackage{amssymb}
\usepackage{color}
\usepackage{float}
\usepackage{amsmath}
\usepackage{amsfonts}
\usepackage{dcolumn}
\usepackage{hyperref}
\usepackage{subfigure}

\usepackage{mathtools}
\usepackage{soul,xcolor}
\hypersetup{
     breaklinks=true,
    pdfstartview={FitH},    
    colorlinks=true,       
    linkcolor=blue,          
    citecolor=red,        
    filecolor=magenta,      
    urlcolor=blue,           
    anchorcolor=green,      
    linktocpage=true
}

\newcommand{\hs}{\hspace{.7mm}}

\begin{document}	
\title{Braneworld inflation with effective $\alpha $-attractor potential}
\author{Nur Jaman}
\email{nurjaman@ctp-jamia.res.in}
\affiliation{Centre for Theoretical
Physics, Jamia Millia Islamia, New Delhi-110025, India}	
\author{Kairat Myrzakulov}
\email{kmyrzakulov@gmail.com}
\affiliation{Center for Theoretical Physics, Eurasian National University, Astana 010008, Kazakhstan}	
\begin{abstract}
 In this paper, we study inflation in $\alpha $-attractor  framework with exponential potential  in the RS braneworlds where high energy corrections to the Friedmann equation facilitate slow roll. In this scenario, we numerically investigate the inflationary parameters and show that the high energy brane corrections have  significant effect on the parameter $\alpha $, namely, the lower values of the parameter are preferred by observation in this limit. The latter substantially reduces the  tensor to scalar ratio of perturbations making the RS brane world inflation compatible with observation.  We also point out that the sub-Planckian values of the field displacement can be achieved by suitably constraining the brane tension.
\end{abstract} 

\maketitle
\section{Introduction}
In the standard framework, a slowly rolling scalar field {\it a la} a shallow field potential may account for inflation\cite{Linde1}. Slow roll along a steep potential is also possible due to Hubble damping caused by high energy brane corrections. Indeed, in brane world scenario, our four dimensional space-time dubbed brane is supposed to be embedded in a higher dimensional bulk \cite{brane,Randall:1999vf} such that the Einstein equations on the brane are modified. The corresponding Friedmann equation includes an extra term\cite{branecor1,branecor2}, quadratic in density which facilitates slow roll in high energy regime at early times even in case of a steep potential\cite{frictionbrane1,frictionbrane2,Gong:2000en,Apostolopoulos:2005ff}. Thus, brane world scenario allows steep potential to support inflation \cite{Sahni:2001qp,Sami:2004xk,Copeland:2000hn} which is not possible in standard case.\\
\indent In the standard cosmology, an exponential potential\cite{expon} does not give  viable inflationary and post inflationary behavior. The situation changes significantly in the brane-world case \cite{Sahni:2001qp,Sami:2004xk,Copeland:2000hn}. However, the steep braneworld inflation gives a ratio of tensor to scalar perturbation, $r$, around $ 0.4$ for 60 $e$-$folds$ of inflation \cite{Copeland:2000hn,Tsujikawa:2003zd} which is not tenable observationally\cite{planck,bicep,Ade:2018gkx,Akrami:2018odb}. Similar problem in the standard cosmology can successfully be addressed in the $\alpha $-attractor scenario\cite{Linde:2016uec,Dimopoulos:2017zvq,Dimopoulos:2017tud,Akrami:2017cir,Shahalam:2018rby,Shahalam:2016juu}. In this framework, the kinetic term in the Lagrangian has a specif non canonical form. 
Canonicalization of such term gives rise to some flat regions or plateaus in the potential\cite{Linde:2016uec,Dimopoulos:2017zvq,Dimopoulos:2017tud,Akrami:2017cir} which are suitable for the study of inflation favored by observational data\cite{planck,bicep,Ade:2018gkx,Akrami:2018odb}. This feature can also be suitable for the study of late time behavior, namely, quintessence \cite{Dimopoulos:2017zvq,Dimopoulos:2017tud,Garcia-Garcia:2018hlc,Peebles:1998qn,samiDE,Hossain:2014zma,Mishra:2017ehw,Sami:2004ic,Akrami:2017cir,Rubio:2017gty}. 

It should be mention here that a super-Planckian displacement of the scalar field may spoil the flatness of quintessential region of the potential and may generate an unwanted fifth force problem\cite{Dimopoulos:2017zvq,Dimopoulos:2017tud,Akrami:2017cir,xxx}. On the other hand, it is impossible to evolve to quintessence starting from the inflationary region without invoking super-Planckian values of the field and not making the potential too curve during inflation\cite{Dimopoulos:2002hm}. The $\alpha$-attractor solves this problem, namely, the canonicalization of the potential makes it possible for the canonical scalar field to have a super-Planckian excursion while keeping its non-canonical counter part under sub-Planckian. \\ 
 In view of the aforesaid, we are led to consider the $\alpha$ attractor construct in the framework of RS brane worlds\cite{Randall:1999vf} which might give new insights related to the  sub-Planckian nature of non-canonical field\footnote{Let us emphasize that in the RS scenario, the   TeV scale associated with electroweak scale has a significance. Since LHC did not see  new excitations around this scale {\it a la} Kluza Klein and no deviations to Newtonian potential were seen at sub-millimeter scale, one could think that the scenario is ruled out.
We should, however, remember, that the requirement of TeV is inspired by the naturalness problem. Let us note that the standard model of electro-weak interactions is also plagued with naturalness problem\citep{Kaul:2008cv}, in that case, we think that there is some  unknown UV completion mechanism required to tackle the issue. Thus in in RS scenario the fundamental scale could well be higher than one TeV and we can still use the scenario using the similar reasoning.}
 
 The structure of this paper is as follows. In section (\ref{SecTheModel}), we discuss how we obtain our effective $\alpha$-attractor potential and suggest some approximations to check for analytical behavior. In section (\ref{SecAlphaOnBrane}), we put the effective potential on brane and  perform a full numerical study of different parameters related to inflation. Numerical analysis is done because of complexity in solving the problem analytically. In this section, we also show some important features our model exhibits. Next, in section(\ref{SecAppAna}) we show some approximated analytical results for our model. Next, in section (\ref{SecConsFromObs}), we compare our results with current observational bounds and with the results obtained, we constrain our different model parameters especially the parameter $\alpha$. Next, in section (\ref{latetime}) we briefly discuss  the late time behavior followed by conclusions in section (\ref{SecConcl}).

\section{The effective $\alpha$-attractor model}\label{SecTheModel}
Considering the formal $\alpha$-attractor  Lagrangian density  with an exponential potential in 4-dimensional space-time\cite{Linde:2016uec,Dimopoulos:2017zvq, Dimopoulos:2017tud,Akrami:2017cir}\footnote{In refs~\cite{Dimopoulos:2017zvq, Dimopoulos:2017tud}, a negative cosmological constant is considered in the Lagrangian to make the vacuum energy density of the universe zero but we do not consider this here as its contribution is insignificant  during inflation.}

\begin{eqnarray}
\mathcal{L}=\frac{\frac{1}{2}\left(\partial \phi\right)^2}{\left(1-\frac{\phi^2}{6\alpha m_p^2}\right)^2} +V_0 e^{-\kappa \phi/m_p} \label{alphalag},
\end{eqnarray}

where $\alpha>0$ is a parameter featuring a pole in the kinetic energy, $m_p=\frac{1}{\sqrt{8 \pi G}}$, is the 4-dimensional reduced Planck mass, $G$ is the Newton's constant, $\kappa$ is the parameter determining the steepness of the potential, $V_0$ is a constant with the dimension of energy density. The modulus value of $\phi$ will remain less than $\sqrt{6\alpha }\,\ m_p$ for any finite value of $\alpha$ because the kinetic energy becomes singular at this value. This allows the  scalar field to remain under sub-Planckian values as long as $\alpha\lesssim1/6$. The same theory cab be described in terms of a canonicalized inflaton field $\varphi$ related to the non-canonical scalar field $\phi$ via the transformation
\begin{eqnarray}
\phi=\sqrt{6 \alpha}\hs m_p\tanh(\frac{\varphi}{\sqrt{6 \alpha}m_p})\label{phiphi}.
\end{eqnarray}
From equation(\ref{phiphi}), it is clear the canonical field $\varphi$ can take any value, keeping the non-canonical $\phi$ sub-Planckian.
By this transformation the potential given in equation(\ref{alphalag}) is described now in terms of the canonical field of the  form
\begin{eqnarray}
V(\varphi)= V_0 e^{-\kappa \sqrt{6 \alpha}\hs \tanh(\frac{\varphi}{\sqrt{6 \alpha}m_p})}\label{V1}.
\end{eqnarray}

\begin{figure}[H]
\centering
  \includegraphics[width=7cm, height=5cm]{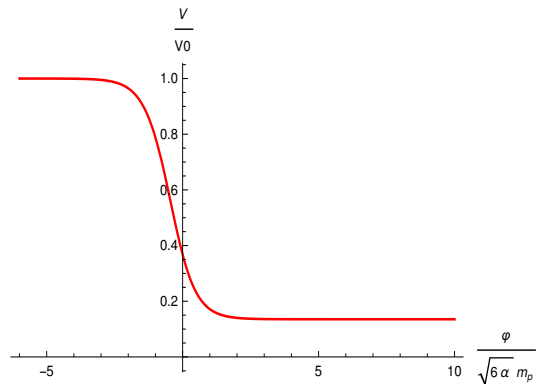}
  \caption{ potential of $\alpha$-attractor after canonicalization (\ref{V1}) \\( $n$ is taken 1)}
  \label{veffpotfig}
\end{figure}

This potential corresponds two plateaus, see Fig.~\ref{veffpotfig}. The inflationary  regime is featured by a plateau corresponds to  $\phi\rightarrow -\sqrt{6\alpha } \,\ m_p$, or equivalently by  $\varphi\rightarrow-\infty$, the other plateau is featured by $\phi\rightarrow \sqrt{6\alpha } \,\ m_p$ or equivalently by $\varphi\rightarrow\infty$, featuring quintessence. For the inflationary limit potential(\ref{V1}) becomes
\begin{eqnarray}
V(\varphi)= M^4 \exp\left( -2 n e^{\frac{2 \sqrt{8 \pi }}{\sqrt{6 \alpha }\hs M_{Pl}}\varphi}\right)\label{veffpot}
\end{eqnarray}
where $M^4=V_0 e^{\kappa \sqrt{6 \alpha}}$, is a constant representing the energy scale for inflation, $M$ has  dimension of mass, $M_{Pl} \equiv m_p \sqrt{8 \pi}$ is the  4-dimensional Planck mass and  $n \equiv  \kappa \sqrt{6 \alpha}$
\section{The effective $\alpha$-attractor potential in braneworld scenario}\label{SecAlphaOnBrane}
\,\ We place our effective potential on the  Randall Sundrum II(RSII) brane\cite{Randall:1999vf} to study the inflationary scenario. The  matter fields are confined to the brane only for RSII model, so our scalar field will remain on the brane only. For a flat Friedmann-Lema\^{\i}tre-Robertson-Walker-Walker(FLRW) background on the brane with zero 4-dimensional cosmological constant, the Friedman equation becomes 
\cite{branecor1,Sahni:2001qp,Sami:2004xk,Tsujikawa:2003zd}
\begin{eqnarray}
H^2\equiv\left(\frac{\dot{a}}{a}\right)^2=\frac{8 \pi}{3 M_{Pl}^2}\rho\left( 1+\frac{\rho}{2\lambda}\right)\label{frdmn1},
\end{eqnarray}

where $a$ is the scale factor, $H$ is the Hubble parameter, $\rho$ is energy density of matter field on the brane, $\lambda$ is 3-brane tension relating the 4-d Planck mass, $M_{Pl}$  with 5-d Planck mass $M_5$ via\\
\begin{eqnarray}
\lambda=\frac{3}{4 \pi}\frac{{M_5}^6} {{M_{Pl}}^2} \label{lamdadef}.
\end{eqnarray}
For high energies, $\rho^2$ term become significant and plays a crucial role in the dynamics of the scalar field, hence of the universe. The scalar field or the inflaton field $\varphi$, confined on the brane  satisfies the Klein-Gordon equation
\begin{eqnarray}
\ddot{\varphi}+3H \dot{\varphi}+V'(\varphi)=0\label{KGeq}.
\end{eqnarray}
$V(\varphi)$ is the potential driving the inflation.  The prime denotes a derivative with respect to $\varphi$ . The presence of quadratic term $\rho^2$  enhances the value of Hubble parameter (\ref{frdmn1}) and hereby gives extra friction to the scalar field (\ref{KGeq}) and makes its evolution slower. Combining  Eqs.~(\ref{frdmn1}) and (\ref{KGeq}) one gets the evolution equation\cite{frictionbrane1,Tsujikawa:2003zd}
\begin{eqnarray}
\frac{\ddot{a}}{a}=\frac{8 \pi}{3 M_{Pl}^2}\left[(V-\dot{\varphi}^2)+\frac{\dot{\varphi}^2 + 2 V}{8 \lambda}\left( 2 V-5 \dot{\varphi}^2\right) \right] \label{roycheq}
\end{eqnarray}
The inflationary condition $\ddot{a} >0$, is reduced to standard form $V>\dot{\varphi}^2 $ \,\ for \,\ $\frac{\dot{\varphi}^2 + 2 V}{8 \lambda}\ll 1$. In the  high energy scenario, the condition becomes $2 V> 5 \dot{\varphi}^2$. This condition may used for characterizing end of inflation \cite{Tsujikawa:2003zd}, $2 V(\varphi_{end}) \simeq 5 \dot{\varphi}_{end}^2 $ .        
Using the slow roll approximation ($V\gg \dot{\varphi}^2$ , $\frac{\ddot{\varphi}}{3 H \dot{ \varphi}}\ll 1$) we can write Eqs.~(\ref{frdmn1}) and (\ref{KGeq}) respectively as
\begin{eqnarray}
&& H^2=\frac{8 \pi}{3 M_{Pl}^2}V\left( 1+\frac{V}{2\lambda}\right)\label{frdmn2}\\
		 \,\,\,\ &&\,\,\,\,\  \mbox{and}\nonumber\\
&&3H\dot{\varphi}+V'(\varphi)=0 \label{KGeq2}.
\end{eqnarray}
These two Eqs.~(\ref{frdmn2}-\ref{KGeq2}) make the condition for inflation end to be 
\begin{eqnarray}
\frac{V^3(\varphi_{end})}{V'^2(\varphi_{end})}\simeq\frac{5 \hs \lambda \hs M_{Pl}^2}{24 \pi}\label{endoinfl}.
\end{eqnarray}
The amplitude of scalar and tensor perturbation in RSII inflationary scenario are given as\cite{frictionbrane1,Tsujikawa:2003zd,Langlois:2000ns,Dinda:2014zta}
\begin{eqnarray}
A_S^2&=& \left.\frac{512}{75 M_{Pl}^6}\frac{V^3}{V'^2}\left(1+\frac{V}{2  \lambda}\right)^3\right\vert_{k=a H}\label{AS},\\
A_T^2&=&\left. \frac{4}{25 \pi}\frac{H^2}{M_{Pl}^2}F^2(x)\right\vert_{k=a H}\label{AT},
\end{eqnarray}
where
\begin{eqnarray}
x&=&H M_{Pl}\sqrt{3/\left(4\pi\lambda\right)} \,\ \simeq \sqrt{\frac{2V}{\lambda}(1+\frac{V}{2\lambda})}\label{x} ,\\
F(x)&=&\left[\sqrt{1+x^2}-x^2\sinh^{-1}(1/x)\right]^{-1/2}\label{F}.
\end{eqnarray}
The '$\simeq$' is used under slow-roll approximation. Amplitudes $A_S $ and $A_T$ are evaluated at horizon exit, $k = a H$, with $k$ being comoving wave number. The two slow roll parameters on the brane are given by
\begin{eqnarray}
\epsilon&\equiv& \frac{M_{Pl}^2}{16\pi}\left(\frac{V'}{V}\right)^2\frac{1+\frac{V}{\lambda}}{\left(1+\frac{V}{2\lambda}\right)^2}\label{epsilon},\\
\eta&\equiv&\frac{M_{Pl}^2}{8\pi}\frac{V''}{V}\frac{1}{1+\frac{V}{2\lambda}}\label{eta}.
\end{eqnarray}
which indicates that in high energy regime($V/\lambda\gg 1$), slow roll is possible even if the potential is steep.
The spectral indices of scalar and tensor perturbations are
\begin{eqnarray}
n_S-1&\equiv&\left.\frac{d \ln{A_S}^2}{d\ln{k}}\right\vert_{k=a H} \label{nsmin1},\\
n_T &\equiv& \left.\frac{d \ln{A_T}^2}{d\ln{k}}\right\vert_{k=a H} \label{nT}.
\end{eqnarray}
Under slow-roll conditions, we get 
\begin{eqnarray}
n_S=1-6\epsilon + 2\eta \label{nsmin1brane}.
\end{eqnarray}
The number of $e$ folds during inflation is given by $\int_{t_*}^{t_{end}} H dt$ , which under slow-roll condition can be written as
\begin{eqnarray}
N\simeq-\frac{8\pi}{M_{Pl}^2}\int_{\varphi_*}^{\phi_{end}}\frac{V}{V'}\left(1+\frac{V}{2\lambda}\right)d\phi\label{efold},
\end{eqnarray}
where $*$ denotes the value at the horizon exit. We define the  ratio of  tensor-to-scalar  perturbation $r$ as \cite{Tsujikawa:2003zd}
\begin{eqnarray}
r\equiv 16\left(\frac{A_T^2}{A_S^2}\right)\label{R1}.
\end{eqnarray}
In the high energy limit $V/\lambda\gg1$ , one finds from Eq.~(\ref{F})
$F^2\simeq \frac{3V}{2\lambda}$, this together with the slow roll approximation ($\rho\sim V$),  and using  Eqs.~(\ref{frdmn2}), (\ref{AS}), and (\ref{AT}) we get,

\begin{eqnarray}
r&=&\frac{M_{Pl}^2}{\pi}\left(\frac{V'}{V}\right)^2\frac{1}{\left(1+V/2\lambda\right)^2}F^2\nonumber\\
&=& \frac{3M_{Pl}^2}{2\pi}\left(\frac{V'}{V}\right)^2\frac{V/\lambda}{\left(1+V/2\lambda\right)^2}\simeq24\epsilon\label{Rbrane}.
\end{eqnarray}
One can easily show that in the low energy limit($V/\lambda\ll1$)$, \,\ r=16\epsilon$, which is the standard expression. 

To study inflation, we start with the potential (\ref{veffpot}). The condition for inflation end (\ref{endoinfl}) gives
\begin{eqnarray}
 && \frac{ 3 \alpha  M^4 M_{Pl}^2}{64 \pi  n^2} \exp \left[-\frac{8 \sqrt{\frac{\pi }{3}} \hs\varphi_{end} }{\sqrt{\alpha } M_{Pl}}-2 n e^{\frac{4 \sqrt{\frac{\pi }{3}}\hs \varphi_{end} }{\sqrt{\alpha } M_{Pl}}}\right]
\nonumber \\
 &\,\,\,\,\ &\simeq 5 \frac{\lambda M_{Pl}^2}{24 \pi}\label{eofmy}
\end{eqnarray}
The total number of $e$-foldings of inflation is given by (\ref{efold}) for high energy limit $(\frac{V}{\lambda}\gg1 $)
\begin{eqnarray}
N \simeq \int_{\varphi_*}^{\varphi_{end}} \frac{\sqrt{3 \pi } \sqrt{\alpha } M^4}{2 n \lambda  M_{Pl} } \exp \left[-\frac{4 \sqrt{\frac{\pi }{3}}\hs \varphi }{\sqrt{\alpha }  M_{Pl}}-2 n e^{\frac{4 \sqrt{\frac{\pi }{3}}\hs \varphi }{\sqrt{\alpha }  M_{Pl}}}\right]d\varphi \nonumber\\
  \,\ \label{Nmy}
\end{eqnarray}
The two slow roll parameters in the high energy limit become
\begin{eqnarray}
\epsilon &\simeq & \frac{16  n^2 \lambda}{3 \alpha  M^4} \exp \left[\frac{8 \sqrt{\frac{\pi }{3}}\hs \varphi }{\sqrt{\alpha }  M_{Pl}}+2 n e^{\frac{4 \sqrt{\frac{\pi }{3}}\hs \varphi }{\sqrt{\alpha } M_{Pl} }}\right]\label{epsilonmy}.\\
\eta &\simeq & \frac{8 n\lambda }{3 \alpha  M^4} \left( 2 n e^{\frac{4 \sqrt{\frac{\pi }{3}}\hs \varphi }{\sqrt{\alpha } M_{Pl}}}-1\right)\exp\left[ \frac{4 \sqrt{\frac{\pi }{3}}\hs \varphi }{\sqrt{\alpha } M_{Pl}}+2 n e^{\frac{4 \sqrt{\frac{\pi }{3}}\hs \varphi }{\sqrt{\alpha } M_{Pl}}}\right]\nonumber\\
\,\ \label{etamy}
\end{eqnarray}
We do solve Eqs. (\ref{eofmy}) and (\ref{Nmy}) numerically and using (\ref{epsilonmy}) and (\ref{etamy}),   we compute $r $ and $n_S$ from the expressions (\ref{Rbrane}) and (\ref{nsmin1brane}).
   
The value of the tensor-to-scalar ratio $r$, is found to be $24/N$, which is $0.4$ for $N=60$ for the standard brane-world scenario\cite{Copeland:2000hn,Tsujikawa:2003zd} without $\alpha-$attractor part, our numerical results show a correction for the tensor-to-scalar ratio. We found that $r$, is depending on the ratio of potential strength $M^4$ to $\lambda$, i.e $\frac{M^4}{\lambda}$ and the parameter $\alpha$, it does not depend on the absolute value of $M^4$ and $\lambda$, which can also be seen in the crude analytical result(see Eq.~(\ref{RAnalytical1})). The following numerical result will confirm the fact. \\
 $r=0.0990187$, for $\frac{M^4}{\lambda}=100,\,\ \alpha=1 ,\,\ \kappa=\sqrt{3}$ for both the value of $M$ equals $0.1 $ and $0.01$ respectively where the value of $\lambda$ is correspondingly chosen.
\begin{figure}
\subfigure[]{
\includegraphics[scale=.9]{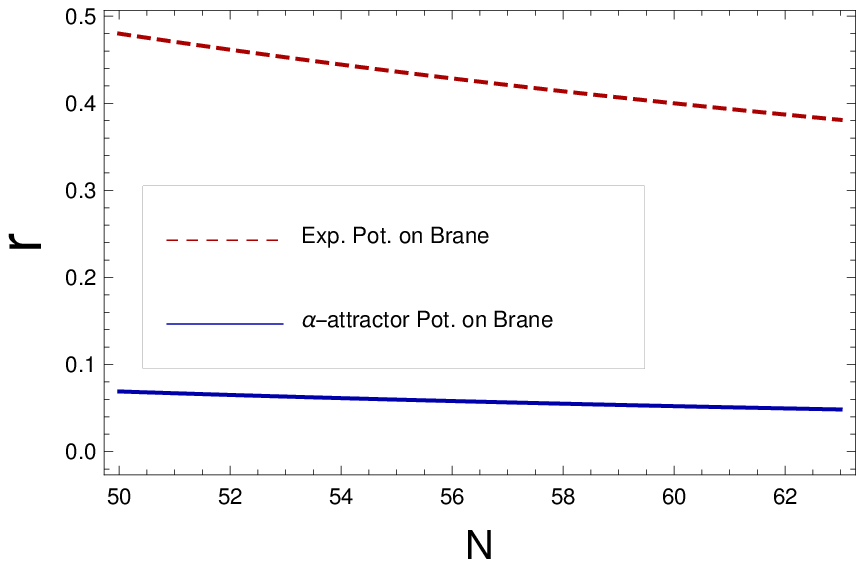}
 \label{figRvsN}}
  \hspace{1 cm} 
\subfigure[]{
 \includegraphics[scale=0.9]{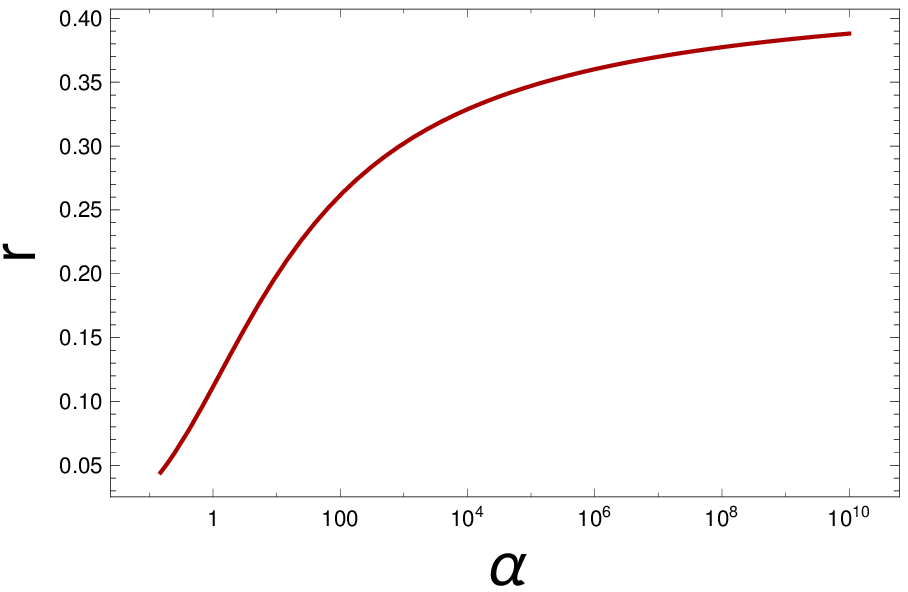}\label{assym}}
\label{fig:test}
\caption{\textbf{Up}: shows the variation of $r$ with $N$, the red (dashed) line for normal exponential potential on brane and the blue line (solid) for the exponential potential on brane with $\alpha$-correction for parameters value $M^4/\lambda=50$ and $\alpha=0.5$. \textbf{Bottom}: shows the asymptotic behavior of $r$ as $\alpha$ increases, for $N=55$ and $M^4/\lambda=100$ }.
\end{figure}
Now we will discuss some important results of our analysis one by one. 
\subparagraph{(\romannumeral 1)  N vs r :}  From the Fig.~\ref{figRvsN}, we see that we have a clear improvement for the value of $r$ those compared to the case standard exponential potential on the brane.
\subparagraph{(\romannumeral 2) asymptotic value for $\alpha$ :} 
In the limit $\alpha\rightarrow \infty$, $\alpha$-attractor correction becomes irrelevant\cite{Linde:2016uec,Dimopoulos:2017zvq}  and we get usual exponential potential. From the Fig.~\ref{assym} it can be seen that $r$ approaches its asymptotic value as we increase the value of the parameter $\alpha$.
\subparagraph{(\romannumeral 3)} It is worth noting that value of $r$ is insensitive to $\kappa$ in the original exponential potential, i.e $n$ for the potential (\ref{veffpot}) which we found to be same from the result we obtained in analytical approximation(\ref{RAnalytical1})
\section{Approximated analytical results}\label{SecAppAna}
An oversimplified  approximation for the potential(\ref{veffpot}) can help us to get an approximate analytical result which we can use as a reference. To do so, we further simplify the potential in the limit $\varphi\rightarrow -\infty$ as\\
\begin{eqnarray}
V\left(\varphi\right)\simeq M^4\left[1-2 n \exp\left(\frac{2 \sqrt{8 \pi}}{\sqrt{6 \alpha}\hs M_{Pl}}\varphi\right)\right].\label{VAnalytical1}
\end{eqnarray}
The condition for end of inflation (\ref{endoinfl}) gives \\
\begin{eqnarray}
\frac{3\hs  \alpha  M^4 M_{Pl}^2 \hs e^{-\frac{8 \sqrt{\frac{\pi }{3}} }{\sqrt{\alpha } M_{Pl}} \varphi_{end}}}{64 \pi  n^2} \simeq \frac{5 \lambda M_{Pl}^2}{24\pi}\label{endofinflana} .
\end{eqnarray}
The Slow roll parameters(\ref{epsilonmy},\ref{etamy}) under this approximation become\\
\begin{eqnarray}
\epsilon & \simeq & \frac{16 \lambda \hs  n^2 \hs e^{\frac{8 \sqrt{\frac{\pi }{3}} \hs \varphi }{\sqrt{\alpha } \hs M_{Pl}}}}{3 \hs \alpha \hs  M^4 \left(1-2 \hs n \hs e^{\frac{4 \sqrt{\frac{\pi }{3}} \varphi }{\sqrt{\alpha } M_{Pl}}}\right)^3}\simeq \frac{16 \lambda \hs  n^2 \hs e^{\frac{8 \sqrt{\frac{\pi }{3}} \hs \varphi}{\sqrt{\alpha } \hs M_{Pl}}}}{3 \hs \alpha \hs  M^4}.\nonumber\\
\label{epsilonana}\\
\eta &\simeq&-\frac{8 \hs \lambda \hs  n \hs  e^{\frac{4 \sqrt{\frac{\pi }{3}} }{\sqrt{\alpha } \hs M_{Pl}} \varphi}}{3 \hs  \alpha \hs  M^4 \left(1-2 \hs n\hs e^{\frac{4 \sqrt{\frac{\pi }{3}} \hs \varphi }{\sqrt{\alpha }\hs  M_{Pl}}}\right)^2}\nonumber\\
&\simeq& -\frac{8 \hs \lambda \hs  n \hs  e^{\frac{4 \sqrt{\frac{\pi }{3}}  }{\sqrt{\alpha } \hs M_{Pl}}\hs \varphi}}{3 \hs \alpha \hs  M^4}\label{etaAnalytical}.
\end{eqnarray}
The amplitude of the scalar perturbation(\ref{AS}) \\
\begin{eqnarray}
A_S^2 &\simeq& \left.\frac{  M^{16}\hs \alpha}{25 \hs \lambda ^3 \hs M_{Pl}^4 \hs n^2} \hs  e^{-\frac{8 \sqrt{\frac{\pi }{3}}\hs  \varphi }{\sqrt{\alpha } \hs M_{Pl}}} \left(1-2 \hs n \hs  e^{\frac{4 \sqrt{\frac{\pi }{3}}\hs  \varphi }{\sqrt{\alpha } \hs M_{Pl}}}\right)^6\right\vert_{k=a H}\nonumber\\
&\simeq& \left.\frac{ M^{16}\hs \alpha \hs  e^{-\frac{8 \sqrt{\frac{\pi }{3}} \hs  \varphi }{\sqrt{\alpha } \hs  M_{Pl}}}}{25 \hs \lambda ^3 \hs  M_{Pl}^4 \hs n^2}\right\vert_{k=a H}.\label{AsqAnalytic}
\end{eqnarray}.
\begin{figure}
  \includegraphics[width=7cm, height=7cm]{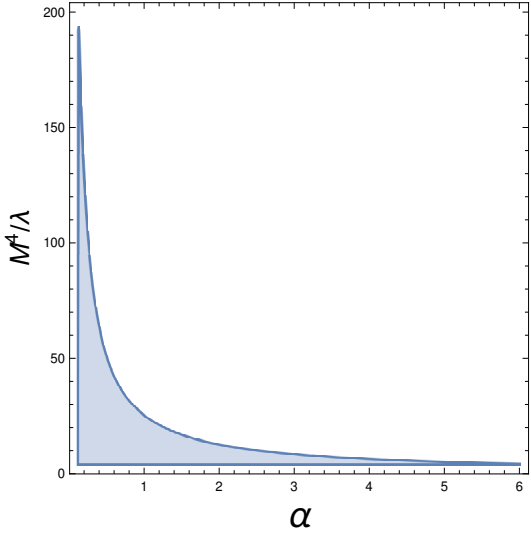}
  \caption{allowed region for $\alpha$ and $M^4/\lambda$ for  $r\leq0.06$}
  \label{regionplt}
\end{figure}

The number of $e$-foldings under this approximation is evaluated to be \\
\begin{eqnarray}
N &\simeq& \int_{\varphi_*}^{\varphi_{end}}\frac{\sqrt{3 \pi } \hs \sqrt{\alpha }\hs  M^4 }{2 \hs  n \hs \lambda  \hs M_{Pl} } \hs e^{-\frac{4 \sqrt{\frac{\pi }{3}}\hs  \varphi }{\sqrt{\alpha } \hs M_{Pl}}} \left(1-2 \hs  n \hs e^{\frac{4 \sqrt{\frac{\pi }{3}}\hs  \varphi }{\sqrt{\alpha } \hs M_{Pl}}}\right)^2 d\phi \nonumber\\
&\simeq& \int_{\varphi_*}^{\varphi_{end}} \frac{\sqrt{3 \pi } \hs \sqrt{\alpha } \hs M^4 \hs  e^{-\frac{4 \sqrt{\frac{\pi }{3}}\hs  \varphi }{\sqrt{\alpha } M_{Pl}}}}{2 \hs  n \lambda \hs  M_{Pl}} d \varphi \,\ .
\label{efoldanalytical}
\end{eqnarray}
Using condition of inflation end(\ref{endofinflana}), we express $N$ in terms of field value at the horizon exit,
\begin{eqnarray}
N\simeq \frac{3 \hs \alpha \hs  M^4 \left(e^{-\frac{4 \sqrt{\frac{\pi }{3}} \hs  \varphi_*}{\sqrt{\alpha }\hs  M_{Pl}}}-\frac{2 \hs  n \hs  \sqrt{10} \hs  \sqrt{\frac{\lambda }{\alpha }}}{3 M^2}\right)}{8 \hs \lambda \hs  n} \label{efoldfinalAna}.
\end{eqnarray}
Using Eq.~(\ref{efoldfinalAna}) and the fact $r=24 \epsilon$, we find the tensor-to-scalar ratio from Eq.~(\ref{epsilonana}), \\
\begin{eqnarray}
r\simeq\frac{288 \hs  \alpha \hs  \lambda \hs  M^4}{\left(\sqrt{10} \hs  \sqrt{\alpha  \lambda } \hs M^2 \hs + \hs  4 \hs N \hs \lambda \right)^2}\simeq\frac{288}{10 \left(1+\frac{4 N}{\sqrt{10 \hs \alpha }\hs  \sqrt{\frac{M^4}{\lambda }}}\right)^2}\nonumber\\
\label{RAnalytical1}
\end{eqnarray}
The amplitude of scalar perturbation(\ref{AsqAnalytic}) is found to be \\
\begin{eqnarray}
A_S^2 &\simeq & \frac{4 M^8 \left(\sqrt{10} \hs  M^2 \sqrt{\alpha  \lambda } \hs + \hs 4 \hs  \lambda \hs  N\right)^2}{225 \alpha \hs  \lambda ^3 \hs  M_{Pl}^4}\nonumber\\
&\simeq& \frac{8 \left(\frac{M^4}{\lambda}\right)^2 M^4}{45 \hs M_{Pl}^4}\left(1+\frac{4 N}{\sqrt{10 \hs  \alpha} \hs \sqrt{\frac{M^4}{\lambda}}}\right)^2.\nonumber\\
\label{AsqAna}
\end{eqnarray}
It should be mentioned here that the relation given by the equations from (\ref{VAnalytical1})to (\ref{AsqAna}) represent only  approximate expressions  for the respective quantities. We  see from (\ref{veffpot}) and (\ref{VAnalytical1}) that this approximation    breaks down for small values of $\varphi$ or  $\alpha < 1$ which is also confirmed by  numerical results.  

\section{Constraining model parameters from observations}
\label{SecConsFromObs}
In order to constrain the parameters of our model we stick to our numerical results. Firstly, we find that $r$ is independent of $M$, $\kappa$ and depends only on the ratio $\frac{M^4}{\lambda}$ and $\alpha$.
\begin{figure}
\subfigure[]{
	\includegraphics[scale=.85]{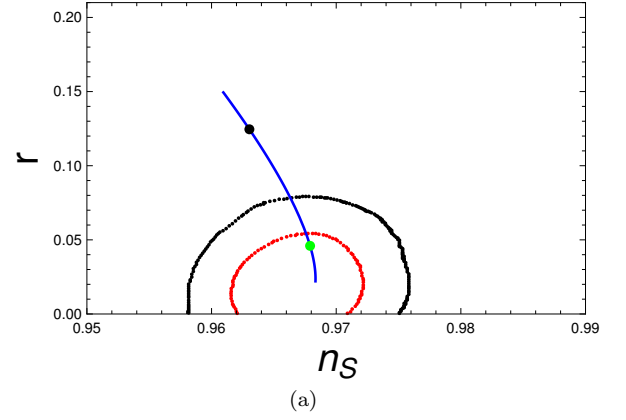}}
  \hspace{1 cm}
\subfigure[]{
  \includegraphics[scale=0.85]{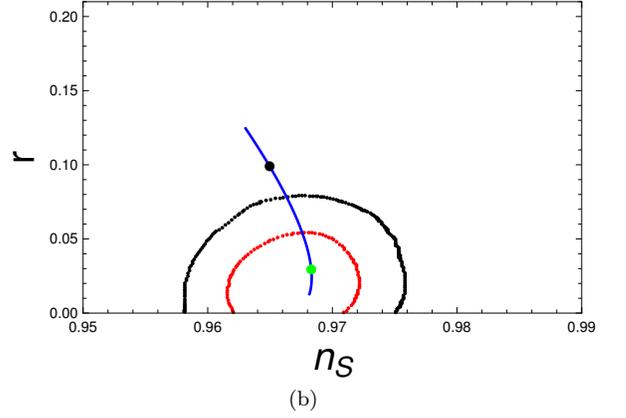} }
\caption{ $68\%$ (red) and $95\%$ (black) contour regions($n_S$-$r$~plane) taken from Planck 2018 results(TT, TE, EE+lowE+lensing+BK14+BAO)\cite{Akrami:2018odb}, we overlay our model's result on it. We obtain our results by varying $\alpha$ from 0.16 to 10. For both the figures black dot and green dot correspond $\alpha=5$ and 0.5 respectively. 
The solid blue line is for $N=60$. For the top Fig., $\frac{M^4}{\lambda}=40$ and for the bottom Fig., $\frac{M^4}{\lambda}=20$.}
\label{cntrplt}
\end{figure}

The observational constraint on the parameter, $r\lesssim 0.06$ \cite{planck,Ade:2018gkx,Akrami:2018odb}, allows us to constrain the parameter $\alpha$ and the ratio $M^4/\lambda$. Theoretically, the high energy limit corresponds $\frac{M^4}{\lambda}\gg1$, but this ratio is highly dependent on  the other parameter $\alpha$ under observational bound. A higher value of the parameter $\alpha $ limits the parameter $M^4/\lambda$ to a lower value. In other words, a decent limit of the assumption that during inflation $M^4/\lambda\gg1$  pushes $\alpha$ to a lower value. The bound $\alpha\leq 39.6$, in the reference \cite{Dimopoulos:2017zvq} is reduced to $\alpha\lesssim 3.6$ for a value of $M^4/\lambda\gtrsim 7$. In Fig.~\ref{regionplt}, we show the allowed  values of $\alpha$ against $M^4/\lambda$. In Fig.~\ref{cntrplt}, we compare our results for different parameters' value with Planck 2018 results\cite{Akrami:2018odb}.

The non-canonical scalar degrees of freedom, $\phi$ remains sub-Planckian as long as $\alpha\lesssim \frac{1}{6}$(\ref{phiphi}). We can obtain this bound in a more compelling way in our model if we consider $\frac{ M^4}{\lambda}\gtrsim 150 $ along with the observational bound  $r\lesssim 0.06$. In other words, the value of $\frac{M^4}{\lambda}$, nearly bigger than $150 $ will always keep the non-canonical scalar degree of freedom within a sub-Planckian value.

\begin{table}[!htb]
\centering
  \begin{tabular}{|p{.8cm}| p{2cm}|}
  \hline
   {$\alpha$} &{ $M$(GeV)} \\
    \hline
   $0.167$ & $1.69 \times 10^{15} $\\ 
   \hline
     $1$&$4.15\times 10^{15}$  \\
    \hline
  
  $10$ & $1.31\times 10^{16}$\\ \hline
 \end{tabular}
 \caption{Values of $M$ for different $\alpha$, $\lambda$ is taken as allowed by $r$ } 
 \label{tableInfscale}  

\end{table}
The COBE normalization corresponds to the amplitude of the scalar perturbations (see Eq.~(\ref{AS})) $A_S\simeq 2\times10^{-5}$\cite{cobe1}, which along with the bound on $\alpha$  determines the energy scale of inflation. We found that (Table-\ref{tableInfscale}) it is near the grand unification scale, almost  same as the one in standard inflationary cosmology. Consequently, after inflation ends, the field will have large overshoot below the background freezing itself for a long time; only at late times it will evolve mimicking cosmological constant like behavior.
\section{Late time behavior}
\label{latetime}
Let us briefly comment on the post inflationary features of the  model. First, the brane corrections to Freidmann equation are insignificant in the post inflationary era. Secondly, it is interesting that irrespective of the nature of original exponential potential, the $\alpha$ attractor effective potential, see Fig.~\ref{veffpotfig} has a generic form, namely, it has plateau followed by a sharp steep behavior like a water fall settling fast to a constant value thereafter. In this case, the tracker\cite{sami:track,Sami:2009jx,Sami:2009dk,Steinhardt:1999nw} behavior is inherently absent which makes the dynamics of scalar field sensitive to its initial conditions. The thawing behaviour in the model under consideration can be understood analytically.
Actually, the important features of dynamics are encoded in a quantity
\begin{figure}
\includegraphics[scale=.8]{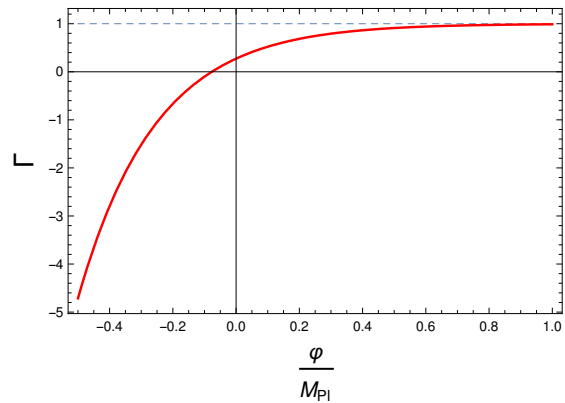}\hspace{0 cm}
\caption{The figure shows the behavior of $\Gamma$ with $\varphi$, for  potential (\ref{V1})  with a residual vacuum energy density $V_0 e^{-\kappa \sqrt{6 \alpha}}$ subtracted from the standard exponential potential in (\ref{alphalag})  as taken in the Refs.~\cite{Dimopoulos:2017zvq,Dimopoulos:2017tud} }, the dashed line represents a constant value 1.
\label{gammaplot}
\end{figure}
dubbed $\Gamma$ which
for the  potential (\ref{V1}) is given by,
\begin{eqnarray}
\label{gammadef}
\Gamma(\varphi)&\equiv& \frac{V(\varphi)''V(\varphi)}{V(\varphi)'^2}\nonumber\\
&=&\frac{e^{-n}}{n}\left\{e^n-e^{n \tanh\left(\frac{2\sqrt{2 \pi/3}}{M_{Pl}\sqrt{\alpha}}\varphi\right)} \right\}\nonumber\\
&&\times\left\{n+\sinh\left(\frac{4\sqrt{2\pi/3}}{M_{Pl}\sqrt{\alpha}}\varphi\right)\right\}
\end{eqnarray}

Eq.~(\ref{gammadef}) tells us that $\Gamma$ increases fast with $\phi$ and crosses zero and approaches unity thereafter mimicking the exponential behaviour, see Fig.\ref{gammaplot}. On the other hand, to realize tracker behavior, it is necessary that $\Gamma$ being greater than unity stays close to one for long time such that the field approximately mimics the background. In the present  case the slope of the potential,starting from a large value, gradually diminishes  pushing the system to slow roll regime at late times.
 Thus owing to the behavior of $\Gamma$ in Fig.~\ref{gammaplot}, the field energy density would witness the large overshoot with respect to the backgrouund in a short span of time freezing the field on its potential due to large Hubble damping. Field evolution would commence only at late stages when background energy density becomes comparable to field energy density allowing the slow roll of field giving rise to late time acceleration; slow roll is characterized 
 by shallow exponential like behaviour.
 Hence, the present scenario gives rise to thawing behaviour as noticed in Ref.\cite{Dimopoulos:2017zvq,Akrami:2017cir} .
\\
\section{Conclusion}
\label{SecConcl}
In this paper, we have considered an inflationary scenario in $\alpha$-attractor 
framework for an exponential potential on RS brane. We have carried out full numerical analysis and presented approximated analytical results. We have found that our results pass the observational constraint for suitable parameter values. The observational bound on the parameter tensor to scalar ratio, $r\leq 0.06$\cite{Ade:2018gkx,Akrami:2018odb} is easily satisfied by our model for a range of parameters$-$ $\alpha\lesssim 3.6$ with $M^4/\lambda\gtrsim 7$. We found that a lower values of the parameter $\alpha$ gives rise to a large  range of the parameter $M^4/\lambda$, falling within the window allowed by observations(Fig.~\ref{regionplt}). The lower bound on $\alpha$, related to the inflation scale,$\alpha\gtrsim10^{-7}$\cite{Dimopoulos:2017zvq}, is not considered here to compare with observational consistency.  The significance of the brane correction underlies with the assumption that $V/\lambda\gg1$ or equivalently $M^4/\lambda\gg1$ during inflation which  automatically pushes $\alpha$ toward a lower values in order to meet observational constraints. We numerically found that for consistency with observation, $M^4/\lambda\gtrsim150$ corresponds to $\alpha <1/6$   which keeps the non-canonical scalar field to be sub-Planckian.  It is worth  mentioning that we do not attempt to constrain here the parameters $V_0$ and $\kappa$ directly as done in \cite{Dimopoulos:2017zvq,Dimopoulos:2017tud}, however, by constraining $\frac{M^4}{\lambda}$ and $\alpha$ puts some indirect constraints on these parameters. In Ref~\cite{Dimopoulos:2017tud} a rather tight bound is given on the parameter $\alpha$ based on the dark energy observations and the super-Planckian issue, $1.5\leq \alpha\leq 4.2$. Our analysis is compatible with this value as we can see from Fig.~\ref{regionplt}. We also find that our inflation is scale is near the grand unification scale, same as the case for standard inflationary models.
As for the post inflationary evolution, we have argued based upon our analytical  expressions that the scenario under consideration should give rise
 to thawing behaviour, see Fig.\ref{gammaplot} and discussion in section V  noticed numerically in Refs.\cite{Dimopoulos:2017zvq,Dimopoulos:2017tud} .
The Present work, with high numerical precision, can be extended  to obtain more accurate bounds on the parameters. The other aspects associated with inflation like reheating can also be investigated for the model under consideration. The investigation of alternative reheating suitable to the the present framework is left for future work.

\section*{Acknowledgments}
N.J is grateful to M. Sami for fruitful discussions. He is also thankful to Bikash and Safia for their help in the work. His work is funded by UGC. We thank Romesh Kaul for fruitful discussions.


\begin{thebibliography}{10}
\bibitem{Linde1} 
 A.D. Linde, Phys. Lett. B,\textbf{129B},177 (1983);  A.~D.~Linde, [arXiv:0705.0164 [hep-th]];

 A.~R.~Liddle,
  astro-ph/9901124.
 
\bibitem{brane} 
  K.~Akama,
  Lect.\ Notes Phys.\  {\bf 176}, 267 (1982)
  [hep-th/0001113]; 
  N.~Arkani-Hamed, S.~Dimopoulos and G.~R.~Dvali,
  Phys.\ Lett.\ B {\bf 429}, 263 (1998)
  [hep-ph/9803315]; 
  L.~Randall and R.~Sundrum,
  Phys.\ Rev.\ Lett.\  {\bf 83}, 3370 (1999)
  [hep-ph/9905221];
  A.~R.~Liddle and A.~N.~Taylor,
  Phys.\ Rev.\ D {\bf 65}, 041301 (2002)
  [astro-ph/0109412].
  
\bibitem{Randall:1999vf} 
  L.~Randall and R.~Sundrum,
  Phys.\ Rev.\ Lett.\  {\bf 83}, 4690 (1999)
  [hep-th/9906064].
\bibitem{branecor1} 
  T.~Shiromizu, K.~i.~Maeda and M.~Sasaki,
  Phys.\ Rev.\ D {\bf 62}, 024012 (2000)
  [gr-qc/9910076].
  \bibitem{branecor2}
  P.~Binetruy, C.~Deffayet, U.~Ellwanger and D.~Langlois,
  Phys.\ Lett.\ B {\bf 477}, 285 (2000)
  [hep-th/9910219]; 
  P.~Binetruy, C.~Deffayet and D.~Langlois,
  Nucl.\ Phys.\ B {\bf 565}, 269 (2000)
  [hep-th/9905012].
\bibitem{frictionbrane1}
  R.~Maartens, D.~Wands, B.~A.~Bassett and I.~Heard,
  Phys.\ Rev.\ D {\bf 62}, 041301 (2000)
  [hep-ph/9912464] .
  \bibitem{frictionbrane2}
   J.~M.~Cline, C.~Grojean and G.~Servant,
  Phys.\ Rev.\ Lett.\  {\bf 83}, 4245 (1999)
  [hep-ph/9906523];
  C.~Csaki, M.~Graesser, C.~F.~Kolda and J.~Terning,
  Phys.\ Lett.\ B {\bf 462}, 34 (1999)
  [hep-ph/9906513]; 
  D.~Ida,
  JHEP {\bf 0009}, 014 (2000)
  [gr-qc/9912002].

\bibitem{Gong:2000en} 
  Y.~g.~Gong,
  gr-qc/0005075.


\bibitem{Apostolopoulos:2005ff} 
  P.~S.~Apostolopoulos, N.~Brouzakis, E.~N.~Saridakis and N.~Tetradis,
  Phys.\ Rev.\ D {\bf 72}, 044013 (2005)
  [hep-th/0502115].


\bibitem{Sahni:2001qp} 
  V.~Sahni, M.~Sami and T.~Souradeep,
  Phys.\ Rev.\ D {\bf 65}, 023518 (2002)
  [gr-qc/0105121].
\bibitem{Sami:2004xk} 
  M.~Sami and V.~Sahni,
  Phys.\ Rev.\ D {\bf 70}, 083513 (2004)
  [hep-th/0402086].

\bibitem{Copeland:2000hn} 
  E.~J.~Copeland, A.~R.~Liddle and J.~E.~Lidsey,
  Phys.\ Rev.\ D {\bf 64}, 023509 (2001)
  [astro-ph/0006421].

\bibitem{expon} F. Lucchin and S. Matarrese
Phys.\ Rev. \ D {\bf 32}, 1316 (1985).
 
\bibitem{Tsujikawa:2003zd} 
  S.~Tsujikawa and A.~R.~Liddle,
  JCAP {\bf 0403}, 001 (2004)
  [astro-ph/0312162].
\bibitem{planck} 
  P.~A.~R.~Ade {\it et al.} [Planck Collaboration],
  Astron.\ Astrophys.\  {\bf 594}, A13 (2016)
  [arXiv:1502.01589 [astro-ph.CO]].
 
 \bibitem{bicep} 
  P.~A.~R.~Ade {\it et al.} [BICEP2 and Keck Array Collaborations],
  Phys.\ Rev.\ Lett.\  {\bf 116}, 031302(2016)[arXiv:1510.09217 [astro-ph.CO]].
\bibitem{Ade:2018gkx} 
  P.~A.~R.~Ade {\it et al.} [BICEP2 and Keck Array Collaborations],
  Phys.\ Rev.\ Lett.\  {\bf 121}, 221301 (2018)
  [arXiv:1810.05216 [astro-ph.CO]].
\bibitem{Akrami:2018odb} 
Y.~Akrami {\it et al.} [Planck Collaboration], arXiv:1807.06211 [astro-ph.CO].

 \bibitem{Linde:2016uec} A.~Linde,
  JCAP {\bf 1702}, no. 02, 028 (2017)
  [arXiv:1612.04505 [hep-th]].

\bibitem{Dimopoulos:2017zvq} 
  K.~Dimopoulos and C.~Owen,
  JCAP {\bf 1706}, no. 06, 027 (2017)
  [arXiv:1703.00305 [gr-qc]].
\bibitem{Dimopoulos:2017tud} 
  K.~Dimopoulos, L.~Donaldson Wood and C.~Owen,
  Phys.\ Rev.\ D {\bf 97}, no. 6, 063525 (2018)
  [arXiv:1712.01760 [astro-ph.CO]].
\bibitem{Akrami:2017cir} 
  Y.~Akrami, R.~Kallosh, A.~Linde and V.~Vardanyan,
  JCAP {\bf 1806}, no. 06, 041 (2018)
  doi:10.1088/1475-7516/2018/06/041
  [arXiv:1712.09693 [hep-th]].
\bibitem{Shahalam:2018rby} 
  M.~Shahalam, M.~Sami and A.~Wang,
  arXiv:1806.05815 [astro-ph.CO].
\bibitem{Shahalam:2016juu} 
  M.~Shahalam, R.~Myrzakulov, S.~Myrzakul and A.~Wang, Int.\ J.\ Mod.\ Phys.\ D {\bf 27}, no. 05, 1850058 (2018) [arXiv:1611.06315 [astro-ph.CO]].
\bibitem{Garcia-Garcia:2018hlc} 
  C.~Garc\'ia-Garc\'ia, E.~V.~Linder, P.~Ru\'iz-Lapuente and Miguel Zumalac\'arregui,
  arXiv:1803.00661 [astro-ph.CO].


  Phys.\ Rev.\ D {\bf 97}, no. 6, 063525 (2018)
  [arXiv:1712.01760 [astro-ph.CO]].
\bibitem{Peebles:1998qn} 
  P.~J.~E.~Peebles and A.~Vilenkin,
  Phys.\ Rev.\ D {\bf 59}, 063505 (1999)
  [astro-ph/9810509].
\bibitem{samiDE} 
  E.~J.~Copeland, M.~Sami and S.~Tsujikawa,
  Int.\ J.\ Mod.\ Phys.\ D {\bf 15}, 1753 (2006)
  [hep-th/0603057].
  \bibitem{Hossain:2014zma} 
  M.~Wali Hossain, R.~Myrzakulov, M.~Sami and E.~N.~Saridakis,
  Int.\ J.\ Mod.\ Phys.\ D {\bf 24}, no. 05, 1530014 (2015)
  [arXiv:1410.6100 [gr-qc]].
\bibitem{Mishra:2017ehw} 
  S.~S.~Mishra, V.~Sahni and Y.~Shtanov,
  JCAP {\bf 1706}, no. 06, 045 (2017)
  [arXiv:1703.03295 [gr-qc]].

  \bibitem{Sami:2004ic} 
  M.~Sami and N.~Dadhich,
  TSPU Bulletin {\bf 44N7}, 25 (2004)
  [hep-th/0405016].
 


\bibitem{Rubio:2017gty} 
  J.~Rubio and C.~Wetterich,
  Phys.\ Rev.\ D {\bf 96}, no. 6, 063509 (2017)
  [arXiv:1705.00552 [gr-qc]].

\bibitem{xxx}
Quintessence: The fifth force - Wetterich, C. Physik J. 3N12 (2004) 43-48 
\bibitem{Dimopoulos:2002hm} 
  K.~Dimopoulos,
  Phys.\ Rev.\ D {\bf 68}, 123506 (2003)
  [astro-ph/0212264].
\bibitem{Kaul:2008cv} 
  R.~K.~Kaul,
  arXiv:0803.0381 [hep-ph].
\bibitem{Langlois:2000ns} 
  D.~Langlois, R.~Maartens and D.~Wands,
  Phys.\ Lett.\ B {\bf 489}, 259 (2000)
  [hep-th/0006007].
\bibitem{Dinda:2014zta} 
  B.~R.~Dinda, S.~Kumar and A.~A.~Sen,
  Phys.\ Rev.\ D {\bf 90}, no. 8, 083515 (2014)
  [arXiv:1404.3683 [astro-ph.CO]].

\bibitem{cobe1} 
  E.~F.~Bunn, A.~R.~Liddle and M.~J.~White,
  Phys.\ Rev.\ D {\bf 54}, no. 10, R5917 (1996)
  [astro-ph/9607038];
  E.~F.~Bunn and M.~J.~White,
  Astrophys.\ J.\  {\bf 480}, 6 (1997)
  [astro-ph/9607060].

\bibitem{Sami:2009jx} 
  M.~Sami,
  Curr.\ Sci.\  {\bf 97}, 887 (2009)
  [arXiv:0904.3445 [hep-th]].
  JCAP {\bf 1706}, no. 06, 045 (2017)
  [arXiv:1703.03295 [gr-qc]].
\bibitem{sami:track}
Models of Dark Energy
M.~Sami, Center forTheoretical Physics, Jamia Millia Islamia, New Delhi, India, \url{https://www.ctp-jamia.res.in/people/models_of_dark_energy.pdf}.

\bibitem{Sami:2009dk} 
  M.~Sami,
  arXiv:0901.0756 [hep-th].
\bibitem{Steinhardt:1999nw} 
  P.~J.~Steinhardt, L.~M.~Wang and I.~Zlatev,
  Phys.\ Rev.\ D {\bf 59}, 123504 (1999)
  [astro-ph/9812313].

\end{thebibliography}
\end{document}